# Vectorization of Hybrid Breadth First Search on the Intel Xeon Phi

Mireya Paredes, Graham Riley and Mikel Luján
School of Computer Science, University of Manchester

## ABSTRACT

The Breadth-First Search (BFS) algorithm is an important building block for graph analysis of large datasets. The BFS parallelisation has been shown to be challenging because of its inherent characteristics, including irregular memory access patterns, data dependencies and workload imbalance, that limit its scalability. We investigate the optimisation and vectorisation of the hybrid BFS (a combination of top-down and bottom-up approaches for BFS) on the Xeon Phi, which has advanced vector processing capabilities. The results show that our new implementation improves by 33%, for a one million vertices graph, compared to the state-of-the-art.

## 1 INTRODUCTION

Modern applications process impressive amounts of data. Graph analysis has emerged as a key area for the analysis of this data as graphs can represent entities in terms of vertices and their relationships in terms of edges. It is common to look for patterns within these relationships, aiming to extract information to be further analysed.

The *Breadth-First Search* (BFS) is one of the main graph search algorithms used for graph analysis and its optimisation has been widely researched using different parallel and distributed systems. From these studies, the BFS parallelisation has been shown to be challenging because of its inherent characteristics, including irregular memory access patterns, data dependencies and workload imbalance, that limit its scalability. However, only 6 papers (see Table 1) have looked at recent parallel architectures using *advanced vector units*; e.g. SIMD Intel AVX-512.

This paper investigates the optimisation and vectorisation of the BFS on the Xeon Phi, which is a parallel architecture containing advanced vector capabilities within the experimental framework of the Graph 500 benchmark. As a result, an novel optimised parallel version of the hybrid BFS is presented using vectorisation, building on top of the vectorised *top-down* BFS introduced by Paredes *et al.* [15]. Note that the concept of the *hybrid* BFS algorithm was introduced by Beamer *et al.* [2]. Our novel hybrid BFS outperforms the state-of-the-art implementation on the Xeon Phi by 33% for one million vertices graphs. This paper presents the vectorisation of the *bottom-up* approach of the hybrid BFS algorithm at first sight is not vector friendly. The vectorisation of the algorithm was implemented by designing an algorithm that restructures the data in a vector friendly manner. For this reason, we believe this paper can be useful for future investigations into new techniques to better exploit vectorisation for irregular data problems. Furthermore, the paper can be a guide for researchers undertaking research on graph algorithms.

The structure of the paper is as follows. Related work is discussed in Section 2. The Xeon Phi architecture is presented in Section 3. We introduce the hybrid BFS algorithm in Section 4. We describe the vectorisation of the bottom-up BFS algorithm in Section 5.1 using the optimization techniques presented in [15]. Section 5 shows an analysis of the bottom-up by using the data taken from hardware performance counters for events related to the instructions, cache access and vector instructions counters. We present performance comparisons between the *non-SIMD* and *SIMD* versions and explore the impact of parallelism at two levels of granularity (OpenMP threading and the vector unit), achieving better results for the native hybrid BFS algorithm on the Xeon Phi than those previously presented in [5] and [6]. The experimental setup and analysis of our results are shown in Section 6. Conclusions and future work are discussed in Section 9.

## 2 RELATED WORK

Table 1 presents a summary that conveys previous studies aiming at the vectorisation of graph algorithms on the Intel Xeon Phi.

| Year | Reference | Approach | optimisation | vectorisation |
|---|---|---|---|---|
| 2012 | Saule and Çatalyürek [17] | top-down | no optimisation | automatic |
| 2013 | Gao *et al.* [19] | top-down | bitmaps | intrinsics |
| 2013 | Stanic *et al.* [18] | top-down | prefetching | intrinsics |
| 2014 | Gao *et al.* [5] | hybrid | bitmaps prefetching | intrinsics |
| 2014 | Golovina *et al.* [6] | hybrid | loop unrolling prefetching | automatic |
| 2015 | Wang *et al.* [20] | hybrid | vertex degree processing | intrinsics |

**Table 1: Previous studies related with the vectorisation of BFS algorithm on the Intel Xeon Phi.**

The key contribution of this work builds on the studies carried out by Gao *et al.* in [19] and [5]. In the first study, they present the vectorisation of the *top-down* BFS algorithm using vector intrinsic functions, which was outperformed by [15], which clarified the impact of *prefetching*, *thread affinity* and vector unit usage rate. The second study is related with the vectorisation of the hybrid BFS algorithm. Similarly to the *top-down* BFS implementation in the first study, Gao *et al.* [19] present the process of vectorising not only the *top-down* but also the *bottom-up* approach of the hybrid BFS algorithm. Again, little detail of their implementation is provided, so in Section 5 we present our vectorisation of the *bottom-up* approach of the hybrid BFS algorithm, including a systematic analysis of the vector unit utilisation. The results of our hybrid BFS algorithm are better in terms of performance compared against those presented in [5].

In addition, the work done by [6] for the vectorisation of the BFS algorithm is listed in the June 2016 Graph 500 list [1]. The experiments were conducted in an Intel Xeon Phi (5110P) platform, similar to the one used in this paper, resulting in 1.80 GTEPS. Section 8 presents a comparison between these results with the results of the vectorisation of the hybrid BFS algorithm implemented in this work.

## 3 THE XEON PHI ARCHITECTURE

The Intel ® Xeon Phi ™ coprocessor used in this paper is composed of 60, 4 way-SMT Pentium-based cores [16] and a shared main memory of 8 GB. Each core contains a powerful 512-bits vector processing unit and a cache memory divided into L1 (32KB) and L2 (512KB) kept fully coherent by a global-distributed tag directory (TD), coordinated by the cache coherency MESI protocol. Cores are interconnected through a high-speed bi-directional ring bus [16] as it is shown in Figure 1. Maximum memory bandwidth is quoted as 320GB/s.

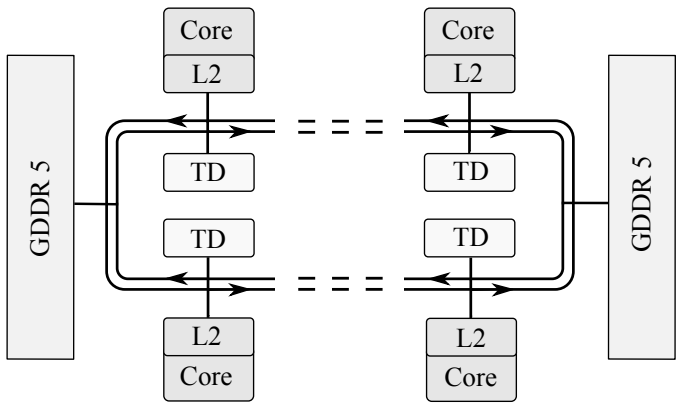

**Figure 1: The Intel ® Xeon Phi Microarchitecture.**

The vector process unit (VPU) is composed of vector registers and 16-bit mask registers. Each vector register can process either 16 (32-bit) operations or 8 (64-bit) operations at a time. A vector mask consists of 16 bits that control the update of the vector elements. Only those elements whose bits are set to 1 are updated into the vector register, the ones with 0 value in the mask remaining unchanged. The Xeon Phi contains both hardware (HWP) and software (SWP) prefetching, which in some cases can help to reduce memory latency.

The programming environment of the Xeon Phi includes multi-threading and the vector units. OpenMP is a versatile interface to program shared-memory multi-threaded systems [4] supported by the Xeon Phi. Programming the vector unit can be done at two levels: automatic and manual. For automatic vectorization, the compiler identifies and optimizes all parts of the code that can be vectorized without the intervention of the programmer. However, there are some obstacles that can limit the vector unit utilization, such as non-contiguous memory accesses or data dependencies [10]. In such cases, manual vectorization can be used allowing the user to force the compiler to vectorize certain parts in the code. Manual vectorization can be set by using SIMD pragma directives supported also using OpenMP library. The compiler also supports a wide range of *intrinsics* which allow a programmer low-level control of the vector unit.

The performance optimisation of a program is a complex task that involves different levels of knowledge [8]. To help this process, there are tools to get access to real-time hardware performance information such as the PAPI (Performance Application Programming Interface) library [14]. This library provides access to various hardware performance counters by tracing different events, while the program is being executed. Five of the PAPI counters were used for the performance analysis of the experiments in Section 7.2.

## 4 THE HYBRID BFS ALGORITHM

The conventional way to traverse the graph is by layers. A *layer* consists of a set of all vertices with the same distance from the source vertex. Processing vertices by layers allows vertices to be explored in any order as long as they are in the same layer; a key feature for further parallelisation. However, each layer has to be processed in sequence. All vertices with distance $k$, (layer $L_k$) are processed before those in layer $L_{k+1}$, which is the reason why the algorithm is often referred to as *layer-synchronous*. The layer-synchronous algorithm uses two lists to represent the concept of a *layer*. The first list contains all the vertices to be processed in the current layer and it will be called **frontier**. The second list is the *output list*, but which for consistency with the literature will also be sometimes referred to as the **output queue**, holds a sequence of vertices that after processing the layer will be swapped with the *frontier* to be processed in the next layer. When a vertex has been processed it is marked as *visited*, otherwise, it remains *non-visited*. Each vertex has an associated set of adjacent vertices to which it is connected by an edge, known as an *adjacency list*. Only the vertices that were found to be *non-visited* vertices are put into the *output queue*. The result of the algorithm is a BFS tree represented by a list of the predecessors or parents ($P$) of the traversed vertices.

Processing vertices in each layer to explore the adjacency list of vertices related to them (parent-child relation) is called *top-down* approach. This is in contrast to the *bottom-up* approach, which is based on processing *non-visited* vertices to find their parents in the frontier (child-parent relation). A combination of both approaches *top-down* and *bottom-up*, called *hybrid*, is the state-of-the-art BFS algorithm [2]. Algorithms 1 and 2 present the top-down and the bottom-up approaches for the exploration per layer of the BFS, respectively, where *in* refers to the input list or frontier, *out* to the output queue, *vis* to the visited array and $P$ to the predecessor array.

---

**Algorithm 1** top-down-BFS($in, out, vis, P$)

1: **for** $u \in in$ **do**
2:    **for** parallel $v \in Adj[u]$ **do**   ▷ Adjacency list exploration
3:       **if** $vis.Test(v) = 0$ **then**   ▷ If $v$ is not visited
4:          $vis.Set(v)$
5:          $out.add(v)$
6:          $P[v] = u$
7:       **end if**
8:    **end for**
9: **end for**

---

**Algorithm 2** bottom-up-BFS($in, out, vis, P$)

1: **for** parallel $v \notin vis$ **do**   ▷ All the *non-visited* vertices
2:    **for** $n \in Adj[v]$ **do**   ▷ Iterating the adjacency list
3:       **if** $n \in in$ **then**   ▷ A parent was found
4:          $vis.Set(v)$
5:          $out.add(v)$
6:          $P[v] = n$
7:          **break**
8:       **end if**
9:    **end for**
10: **end for**

---

Whether a layer is processed using the *top-down* or *bottom-up* approach is dictated by an *online* heuristic based on three metrics; the number of edges to check in the frontier, $e_f$, the number of vertices in the frontier, $v_f$, and the number of edges to check in the *non-visited* vertices, $e_u$. These are variables which are updated during the traversal of each layer. The $e_f$ adds up the degree of every node in the layer, $v_f$ sums the number of vertices added to the layer and $v_u$ counts the number of edges of the *non-visited* vertices. The heuristic compares these metrics at the end of each layer and the resulting information is used to determine whether the layer should be processed using the *top-down* or the *bottom-up* approach.

The pseudocode of the hybrid BFS algorithm is presented in Algorithm 3, showing the switching points to transition between the *top-down* and the *bottom-up* algorithms. Particularly, the vectorisation of the hybrid involves the vectorised version of both algorithms. The vectorisation of the *top-down* algorithm is described and analysed in [15], whereas the vectorisation of the *bottom-up* is described further in Section 5.

Table 2 shows an example of the switching points to swap between the *top-down* and the *bottom-up* approaches of the

---

**Algorithm 3** hybrid-BFS($G, s$)

1: topdown=true
2: **while** $in \neq 0$ **do**
3:    **if** $|in| > f(n, e_f, v_f, e_u)$ **then**   ▷ function f is architecture specific
4:       topdown = false
5:    **else if** $|in| < g(n, e_f, v_f, e_u)$ **then**   ▷ function g is architecture specific
6:       topdown = true
7:    **end if**
8:    **if** topdown **then**
9:       top-down-BFS(in, vis, out, P)
10:    **else**
11:       bottom-up-BFS(in, vis, out, P)
12:    **end if**
13:    $e_f, v_f, e_u, \leftarrow$ getCounters()
14:    swap(in, out)
15:    $out \leftarrow 0$
16: **end while**

---

hybrid BFS algorithm for a graph created by the Graph 500 graph generator introduced in Section 6. The graph size is defined by two parameters: *SCALE* and *edgefactor*. The total number of vertices in the graph is calculated by $2^{SCALE}$ and the total number of edges is calculated by $2^{SCALE} * edgefactor$.

| Layers | $v_f$ | $e_u$ | $f$ | $g$ | Approach |
|---|---|---|---|---|---|
| 1 | 1 | 262,143 | 255 | 4,096 | top-down |
| **2** | **554** | 261,589 | **255** | 4,096 | **bottom-up** |
| 3 | 97,725 | 163,864 | 160 | 4,096 | bottom-up |
| 4 | 77,711 | 86,153 | 84 | 4,096 | bottom-up |
| **5** | **868** | 85,285 | 83 | **4,096** | **top-down** |
| 6 | 5 | 85,280 | 83 | 4,096 | top-down |
| | | | | **TEPS** | 1,009,411,193 |

**Table 2: Example execution of the hybrid BFS for a graph size SCALE=18 and edgefactor = 16.**

## 5 THE BOTTOM-UP VECTORISATION

The *bottom-up* approach consists of stepping through each of the *non-visited* vertices in a layer to find the first vertex in their adjacency list in the frontier. In general, the vectorisation of an algorithm can be tricky in the sense that input data need to be structured in a way that increases vector unit utilisation. Specifically, the vectorisation of the *bottom-up* algorithm does not follow the same principle of vectorising the adjacency list as is used in the *top-down*, which consists of processing the adjacency list in chunks of 16 elements. This is mainly because while the *top-down* approach intends to process the largest number of vertices at the same time, the *bottom-up* aims only to find one parent of each vertex to be processed. For that reason, the vectorisation of the *bottom-up* involves an extra step to gather the input vertices in a layout that better utilises the vector unit, allowing to set multiple parents at a time. This algorithm is presented next.

## 5.1 Setting Multiple Parents

While the *bottom-up* in Algorithm 2 looks to set one parent of each *non-visited* vertex at a time, the vectorised algorithm looks to set parents for multiple vertices at a time. The algorithm consists of the following four steps to process 16 input vertices at a time aiming to find a parent for each by iterating through their adjacency lists.

(1) **Load input vertices.** A sequence of 16 consecutive vertices is loaded to a vector register in a loop over all vertices, starting with vertex 0.
(2) **Filtering *non-visited* vertices**. The input vertices are filtered, so that the only vertices being processed are the *non-visited* vertices. This filtering can be done by using a bit vector mask formed from the visited bitmap array.
(3) **Adjacent vertices iterations**. Each of the adjacent vertices of the *non-visited* vertices are gathered over a number of iterations. In the first iteration, the first vertex in the adjacency list for each *non-visited* vertex is gathered. Subsequent iterations gather the next adjacent vertex. The number of iterations is determined by a statically set threshold related with the average number of iterations required before a parent is found. Figure 2 illustrates the gathering of the adjacent vertices of the $n^{th}$ iteration. After the adjacent vertices are gathered, they are tested to verify if they have been visited previously. The adjacent vertices that are in the frontier of the layer will be set as the parents of their respective vertex and they are no longer considered for processing in the next iteration. After the threshold is reached, the vectorised version is no longer efficient and the algorithm swaps to the non-SIMD version.
(4) **Execute the *non-SIMD* version**. After the threshold in the adjacent vertices is achieved, the input vertices that have not yet found a parent are processed by the using the *non-SIMD bottom-up* algorithm introduced in Algorithm 2.

Algorithm 4 shows the pseudocode of the vectorisation of the *bottom-up* BFS algorithm. The process starts by loading a sequence of 16 consecutive vertices to be processed, starting with vertex zero up to the total number of vertices in the graph. Thus, the vertices are filtered so only the *non-visited* vertices are going to be processed. Then, to do so the visited bitmap array is iterated through stepping every word (32-bits integer) as it is shown in line 1. Since each word is 32 bits in length, but only 16 (32-bit) elements can be loaded into the vector register, the word is processed in two halves as shown in lines 3 to 5. The function *LoadVertices* transforms a single half word into a vector of 16 vertices, with the word parameter specifying whether the upper or lower half word should be processed. Function *GetHalf* loads the 16 bits of a half word into the bit mask $mask_{vis}$. This bitmask is used further to filter the *visited* vertices out.

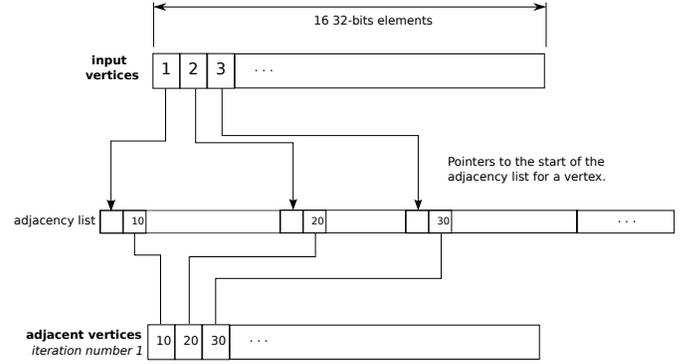

**Figure 2: Example of gathering the adjacent vertices for the iteration number 1 of the search through the adjacency list for specific input vertices.**

**Algorithm 4** bottom-up-multiple-set($in, vis, out, P$)

```
1: for u ∈ vis do            ▷ Stepping through every word (32-bits) in vis.
2:     word = 0
3:     while word < 2 do
4:         vertices ← LoadVertices(u, word)   ▷ Loading 16 vertices.
5:         mask_vis ← GetHalf(u, word)
6:         word = word + 1
7:         pos = 0
8:         for pos < MAX_POS do
9:             LookingParents(in, vis, out, P, vertices, pos, word,
10:                mask_vis, mask)
11:            pos = pos + 1
12:        end for
13:        i = 0
14:        while mask.getBit(i) == 0 AND i < 16 do  ▷ Every bit in flag.
15:            bottom-up-noSIMD()
16:            i++
17:        end while
18:    end while
19: end for
```

The process continues by iterating through the adjacent vertices of each vertex in the input vertices. The maximum number of iterations is delimited by a threshold called $MAX\_POS$. This threshold is a constant calculated based on the minimum number of iterations within which a parent is most likely to be found. In each iteration the function $LookingParents()$ is called to process the 16 input vertices simultaneously aiming to find their parents. If no parent vertex is found up to the threshold, the search continues using the *non-SIMD bottom-up* algorithm presented in Algorithm 2.

## 5.2 Setting the threshold $MAX\_POS$

The threshold that determines the number of iterations to be processed is related to the average number of iterations through the adjacency list before a parent is found. Table 3 shows the average number of edges processed per visited vertex per layer during the 64 iterations of the *bottom-up* BFS algorithm. Notice that the average was calculated by taking into account only the starting vertices that lead to connected subgraphs, unconnected starting vertices were ignored. Also, during the 64 iterations, the number of layers varied between 5 and 6 for different starting vertices, hence the average for the sixth layer was calculated based on only the non-zero values. The results in Table 3 show that the middle layers (layers 3, 4 and 5) have smaller averages of 34.0, 1.06 and 2.51, respectively, implying that the *bottom-up* BFS algorithm is more effective for the middle layers. However, the average for the third layer is above the average for the fourth and the fifth layers. Due to the high variability of the number of edges per visited vertex for the third layer, where most of the cases is under 8, the value of the $MAX\_POS$ threshold used for the further experiments of the *bottom-up* BFS algorithm is statically set to 8.

| Layer | Avg Edges/Visited |
|---|---|
| 1 | 2,879,193 |
| 2 | 30,910 |
| 3 | 34 |
| 4 | 1.06 |
| 5 | 2.51 |
| 6 | 41.06 |

**Table 3: Average of the number of edges processed per visited vertex using the the *bottom-up* BFS algorithm per layer. The graph size is $SCALE = 18$ and the $edgefactor = 16$.**

Algorithm 5 shows the pseudocode of the *LookingParents* function used for processing the adjacent vertices.

(1) Firstly the 16 adjacent vertices of the "input vertices" are loaded into the vector register using the *LoadAdj* function (line 1). This function gathers the $n^{th}$ adjacent vertices of the input *vertices*, only the ones set to zero in the bitmask $mask_{vis}$ are candidates to be processed. The *starts* and *ends* input lists are the starting and the ending indexes in the adjacency list of each vertex and *pos* is the $n^{th}$ location within this range. If *pos* exceeds this range, a mask ($mask_{pos}$) is set so as to not take into account that vertex for further processing in the next iteration. The return value is a vector that holds a list of 16 adjacent vertices.

(2) Secondly, the *frontier* of the adjacent vertices is gathered by the function *Gather* (line 2). This function gathers the respective values in the *frontier* of the current adjacent vertices in the vector *vadj*, only for the vertices set to zero in the input mask $mask_{pos}$. The result of this function is a 16-bits mask, in which each bit indicates whether an adjacent vertex in *vadj* is in the *frontier* or not.

(3) Thirdly, the bitmask is tested to verify whether at least one of the adjacent vertices has been found to be in the frontier of the current layer (line 3).

(4) Finally, the parents found are scattered back to the predecessor array (P) and the visited and output queue bitmap arrays. Additionally, a mask, which is received as input parameter of the function, is updated each time new parents have been found. This mask is used in further iterations to identify the vertices that have already had a parent found to prevent them from being processed in further iterations.

Listing 1 shows the source code of the `LookingParents` function. The half-word of the input visited array is passed as parameter ($mask_{vis}$) to filter out the *visited* vertices, so only the *non-visited* vertices are loading as in line 1. The frontier (*in*) of the adjacent vertices is gathered and test, so only the vertices that are set to one in the frontier are the ones to be updated (lines 2 and 3). Hence, the predecessor array, P, the visited and the output queue bitmap arrays are updated. Additionally, a mask (input parameter) is updated every time new parents have found.

**Algorithm 5** LookingParents($in, vis, out, P, vertices, pos, mask_{vis}, mask_{pos}$)

1: vadj ← LoadAdj(vertices, starts, ends, pos, $mask_{pos}, mask_{vis}$)
2: frontier ← in.Gather(vadj, $mask_{pos}$)
3: **if** $frontier.Test()! = 0x0000$ **then**     ▷ At least there is one parent in the frontier.
4:     P.Scatter(vertices, vadj, frontier)
5:     vis ← vis ∪ frontier
6:     out ← out ∪ frontier
7:     mask ← mask ∪ frontier
8: **end if**

## 6 EXPERIMENTAL SETUP

### 6.1 Hardware platform

First we evaluate the *non-SIMD* presented in Algorithm 2 and the vectorized version, *SIMD*, of the bottom-up BFS algorithm on the Intel Xeon Phi, also known as Knights Corner (5110P). After this step, we evaluate the overall performance of our hybrid BFS. We use OpenMP as a thread parallel platform and IntelAVX-512 intrinsics for the vectorization. We compiled our code with the Intel C++ compiler (version 14.0.0) with the optimization flag *-O2* that allows to execute manual vectorisation using intrinsic functions and *-fopenmp* to link the OpenMP library.

### 6.2 Input graph and Settings

Our implementation uses different modules of the Graph500 benchmark [21], including the graph generator, the BFS path validator, the experimental design and the ability to run our parallel BFS implementation. Firstly, the graph generator creates synthetic scalable large *Kronecker* graphs [12] and is based on the R-MAT random graph model [3]. These graphs aim to naturally generate graphs with common real network properties in order to be able to analyse them. The graph size is defined by the *SCALE* and the *edgefactor* values. The total number of vertices in the graph is calculated by $2^{SCALE}$ and the total number of edges generated by $2^{SCALE} * edgefactor * 2$ (the factor of 2 reflects the fact that

```
__inline void SBFS_2QBM_hybrid_SIMD::LookingParents(int vis_mask,
bitmap_t *frontier, bitmap_t *queue, bitmap_t *explored, int pword,
int psegment, int &fend, int pos,__m512i vstart, __m512i vend, __m512i\
vvertices){
  __m512i vtmp = _mm512_set1_epi32(pos);
  __m512i vadd = _mm512_add_epi32(vstart, vtmp);
  __mmask16 vcmp = _mm512_cmpgt_epi32_mask(vend, vadd);
  __m512i vneig = _mm512_mask_i32gather_epi32(_mm512_set1_epi32(0), vcmp,
    vadd, rows, sizeof(word_t));
  int res = 0;
  //Getting the high part of the word (16-bits)
  vis_mask = (explored->start[pword]>>(psegment*VNELE8))&0xFFFF;

  __mmask16 mask1 =  _mm512_kand(_mm512_knot(_mm512_int2mask(vis_mask)),vcmp);

  //2.- filter visited adjacent vertices according to vis and out

  //Getting WORD offset and BIT offset
  __m512i vword  = _mm512_srlv_epi32(vneig, _mm512_set1_epi32(5));

  _mm512_set1_epi32(BITS_PER_WORD));
  __m512i vbits = _mm512_and_epi32(vneig,_mm512_set1_epi32(0x1F));

  __m512i fron_words = _mm512_mask_i32gather_epi32(_mm512_set1_epi32(0), mask1,
    vword, frontier->start, sizeof(word_t));

  //Shifting 1 to the left indexes position in the vneig array
  __m512i bits= _mm512_sllv_epi32(_mm512_set1_epi32(1), vbits);

  //Filtering with the unvisited mask
  __mmask16 mask = _mm512_mask_test_epi32_mask(mask1 ,fron_words, bits);

  res = _mm512_mask2int(mask);

  if(mask != 0x0000 ){ //at least one neighbour is in the frontier
    _mm512_mask_prefetch_i32scatter_ps(bfs_tree, mask, vvertices,sizeof(word_t),
    _MM_HINT_T0);
    _mm512_mask_i32scatter_epi32(bfs_tree, mask, vvertices, vneig, sizeof(word_t));
    explored->start[pword] |= res<<(psegment*VNELE8);
    queue->start[pword] |= res<<(psegment*VNELE8);
  }

  //flag used to indicate the end of the processing
  fend |= (res | ((explored->start[pword]>>(psegment*VNELE8))&0xFFFF));
}
```

**Listing 1: Source code of the *LookingParents()* SIMD function.**

the edges are bidirectional). The generator uses four initiator parameters (*A, B, C* and *D*), which are the probabilities that determine how edges are distributed in the adjacency matrix representing the graph. We used the standard set of parameters defined by Graph500 (A=0.57, B=0.19, C=0.19 and D=0.05).

The evaluation consists of comparing the performance in terms of number of TEPS, which is a *Graph500* performance metric amply used by other BFS implementations in different architectures to compare them. The input parameters for the experiments are for graph sizes of SCALE (18,19 and 20) and edgefactor (16, 32 and 64), a number of threads of 228 with balanced affinity.

### 6.3 Implementation details

Firstly, the input graph is efficiently represented by a Compressed Sparse Row (CSR) matrix format. Secondly, the experimental design comprises of 64 BFS executions by randomly choosing different starting vertices. For each execution we measure its execution time. After the completion of those executions some statistics related with time and Traversed Edges Per Second (TEPS) are collected. During the experiments we found out that out of the 64 graph500 iterations, some of the starting points are unconnected, which turns out into zero TEPS results in those iterations. This ends up in having a harmonic mean, calculated by Graph500, higher than the maximum number of TEPS. Our results show harmonic mean of the TEPS across the 64 executions to compare with other ranked systems such as [5] and [2].

## 7 BOTTOM-UP ANALYSIS

The previous performance analysis of the vectorisation of the *bottom-up* algorithm is one of the contributions of this work and it consists of a systematic comparison of two versions of the *bottom-up*, the non-SIMD and the SIMD. This comparison is based not only in terms of performance (TEPS) but also in terms of the effectiveness of the utilisation of resources of the Xeon Phi architecture, including counting instructions and cache access. Firstly, to be able to compare the PAPI events for the function calls related to the *bottom-up* algorithm, a specific starting vertex was selected by using the criterion explained below. Secondly, the PAPI events for the instructions counters and the cache access are gathered for the two *bottom-up* implementations and analysed. Finally, a performance analysis in terms of TEPS is presented in Section 8.

### 7.1 Selecting a starting vertex

Each one of the 64 executions computes the BFS with a randomly chosen starting vertex. Despite the 64 starting vertices being selected by a random process, they remain the same across different executions of the experiment as each experiment is generated by the same seed [13].

This variability between iterations is due to the parallel execution of BFS being non deterministic [11], which implies that different executions with the same starting vertex can lead to different valid output BFS trees.

For this reason, based on experiments as means to control the variation, the following results for the *bottom-up* analysis focus on the fourth execution as it presents low variation from run-to-run.

### 7.2 PAPI Counters Analysis

The two *bottom-up* implementations: the non-SIMD and the SIMD version were instrumented to have direct access to some of the hardware events of the Xeon Phi processor using the PAPI library. By monitoring these events, it is possible to use the performance counters to help analyse each of the two implementations in terms of the usage of the resources on the Xeon Phi architecture. PAPI counters are turned on and off around the code segments that are to be analysed. Despite the Xeon Phi having several performance counters, in this analysis five of them were selected to monitor the instructions of the program, cache memory accesses and the vector unit. The counters are divided in two sets, the instructions counters and those related to cache access. First, the instructions counters include the total number of *cycles* and the total number of *instructions*, and the CPI (cycles per instruction) is calculated based on them, Tables 4 and 5. The CPI can be seen as a metric to measure performance consisting of calculating the average number of clock cycles per instruction for a program [7]. Second, the cache access counters include the *L1 data cache misses* and the *L2 load misses*. In addition, an extra hardware counter relevant to analyse the vector unit utilisation is the number of *vector or SIMD instructions*, Tables 6 and 7. The analysis of the five counters is presented next. To have a clear analysis of the data, the experiments in the following sections were conducted by executing only one thread. Tables 4, 5 list the layers of the hybrid algorithm, followed by the approach used to traverse each layer. Specifically, the highlighted rows in grey are the layers processed by the *bottom-up* and those are the ones to focus on this analysis. Additionally, the results show the number of *non-visited* vertices in each layer which is relevant to the *bottom-up* approach. Finally, the performance counters for cycles and instructions are displayed as well as the CPI calculation.

Based on the results, we can see whether in all cases the middle layers (3-5) use the non-SIMD or a vectorised/SIMD version of the *bottom-up*. Despite the fact that there is only a single layer where the *bottom-up* approach is found to be applied, the impact on this layer is beneficial to the final performance.

| Layer | Approach | NV | Time(sec) | Cycles | Inst. | CPI(cyc/inst) |
|---|---|---|---|---|---|---|
| 1 | TD-SIMD | 262,143 | 0.000092 | 105,945 | 15,772 | 6.72 |
| 2 | TD-SIMD | 262,142 | 0.000458 | 471,431 | 84,169 | 5.60 |
| 3 | BU-noSIMD | 259,153 | 0.000844 | 903,366 | 175,435 | 5.15 |
| 4 | BU-noSIMD | 96,758 | 0.000271 | 295,299 | 43,854 | 6.73 |
| 5 | BU-noSIMD | 62,373 | 0.000128 | 140,088 | 26,673 | 5.25 |
| 6 | TD-SIMD | 62,295 | 0.000086 | 96,070 | 14,091 | 6.82 |

**Table 4:** Hybrid BFS execution for starting vertex $12,119$ using the *bottom-up* non-SIMD version. PAPI was used to get the hardware counters for cycles, instructions and CPI(cycles/instructions). $SCALE = 18$ and $edgefactor = 32$.

| Layer | Approach | NV | Time(sec) | Cycles | Inst. | CPI(cyc/inst) |
|---|---|---|---|---|---|---|
| 1 | TD-SIMD | 262,143 | 0.000085 | 96,417 | 14,269 | 6.76 |
| 2 | TD-SIMD | 262,142 | 0.00045 | 462,515 | 81,132 | 5.70 |
| 3 | BU-SIMD | 259,153 | 0.00063 | 687,282 | 99,539 | 6.90 |
| 4 | BU-noSIMD | 96,758 | 0.000272 | 302,230 | 40,385 | 7.48 |
| 5 | BU-noSIMD | 62,373 | 0.000123 | 135,862 | 25,139 | 5.40 |
| 6 | TD-SIMD | 62,295 | 0.000086 | 94,236 | 12,818 | 7.35 |

**Table 5:** Hybrid BFS execution for starting vertex $12,119$ using the *bottom-up* BFS vectorised version (SIMD). PAPI was used to get the hardware counters for cycles, instructions and CPI(cycles/instructions). $SCALE = 18$ and $edgefactor = 32$.

| Layer | Approach | NV | Time(sec) | L1 misses | L2 misses | Vec. Inst. |
|---|---|---|---|---|---|---|
| 1 | TD-SIMD | 262,143 | 0.000088 | 212 | 210 | 45 |
| 2 | TD-SIMD | 262,142 | 0.000448 | 323 | 189 | 181 |
| 3 | BU-noSIMD | 259,153 | 0.000881 | 1,877 | 2,168 | 24 |
| 4 | BU-noSIMD | 96,758 | 0.000261 | 774 | 752 | 24 |
| 5 | BU-noSIMD | 62,373 | 0.000126 | 435 | 151 | 24 |
| 6 | TD-SIMD | 62,295 | 0.000083 | 204 | 118 | 45 |

**Table 6:** Hybrid BFS execution for starting vertex $12,119$ using the *bottom-up non-SIMD* version. PAPI was used to get the hardware counters for L1 and L2 cache misses and vector instructions. The graph size is $SCALE = 18$ and $edgefactor = 32$.

| Layer | Approach | NV | Time(sec) | L1 misses | L2 misses | Vec. Inst. |
|---|---|---|---|---|---|---|
| 1 | TD-SIMD | 262,143 | 0.000089 | 216 | 183 | 45 |
| 2 | TD-SIMD | 262,142 | 0.000458 | 350 | 271 | 181 |
| 3 | BU-SIMD | 259,153 | 0.000627 | 2,329 | 2,828 | 18,099 |
| 4 | BU-noSIMD | 96,758 | 0.000273 | 793 | 870 | 24 |
| 5 | BU-noSIMD | 62,373 | 0.000124 | 423 | 224 | 24 |
| 6 | TD-SIMD | 62,295 | 0.000086 | 235 | 215 | 45 |

**Table 7:** Hybrid BFS execution for starting vertex $12,119$ using the *bottom-up* vectorising by layers algorithm. PAPI was used to get the hardware counters for L1 and L2 cache misses and vector instructions. The graph size is $SCALE = 18$ and $edgefactor = 32$.

Focusing first on the PAPI instruction counters, we can observe in the third layer in Table 5 that the *cycles* and *instructions* are better than in Table 4, even though the *CPI* is worse in Table 5. The cycles count for the non-SIMD and the SIMD are in the same proportion as the time, as expected, since the clock rate of the Xeon Phi is the same. Comparing the number of instructions, in the SIMD version is the number is lower than that for the non-SIMD but the CPI for SIMD is higher than that for the non-SIMD, suggesting that the instructions executed for SIMD are not as efficient as for non-SIMD version. This may be caused as a result of the SIMD version using vector instructions.

Focusing on the PAPI Cache Access and Vector Instructions Counters, we can identify that the cache accesses counters (labeled L1 and L2 misses) are generally lower for the *non-SIMD* version than the vectorised version (BU-SIMD). For instance, for the third layer of the *non-SIMD* version in Table 6, the L1 and L2 misses are 1,877 and 2,168 respectively, which are lower than the ones from the vectorised version either using the *SIMD* (2,329) in Tables 7.

There is also a strong correlation between the cache access and the vector instructions counters. Comparing the vector instructions counters between the *non-SIMD* version in Table 6 and the counters for the vectorised version, in Table 7, the number of vector instructions is much higher using BU-SIMD version and remains constant for the layers using the BU-noSIMD version. For those layers using the BU-SIMD version, the number of L1 and L2 misses is higher than the ones using the BU-noSIMD version.

## 8 HYBRID BFS ANALYSIS

Previous section has compared the *non-SIMD* and *SIMD* versions of the *bottom-up* BFS implementations and thus we have shown the benefit of using vectorisation for the *bottom-up* approach.

In this section a second set of results compare our hybrid BFS algorithm using vectorisation against the state-of-the-art hybrid BFS described by Gao *et al.* [5] targeting the Xeon Phi.

*No-SIMD versus SIMD.* Figure 3 shows the results of the hybrid BFS algorithm using the *non-SIMD* version and the *SIMD bottom-up* versions for different graph sizes varying the *SCALE* from 14 up to 22 and for *edgefactor* 16, 32 and 64 (a, b and c in the figure).

(1) For the three cases in Figure 3 (a, b and c), the performance of the SIMD version (red line) is higher than the *non-SIMD* (blue line), as was expected as a result of exploiting the vector unit. The speedup is approximated around 250 MTEPS (7.14%), 500 MTEPS (9.09%) and 790 MTEPS (8.98%) respectively. Moreover, the higher the *edgefactor*, the wider is the gap between both versions, keeping a similar shape as long as the graph size increases. The reason for this is that the vectorisation for the *bottom-up* algorithm does not apply the same strategy for vectorising the adjacency list as the *top-down* does, which is affected by the edgefactor. Instead, the *bottom-up* sets up a threshold, introduced in Section 5.1, to only iterate through certain number of adjacent vertices. Thus, even though the *edgefactor* might be bigger, which means that the number of adjacent vertices is larger, the *bottom-up* cuts off iterating through all the adjacent vertices due to this static threshold.

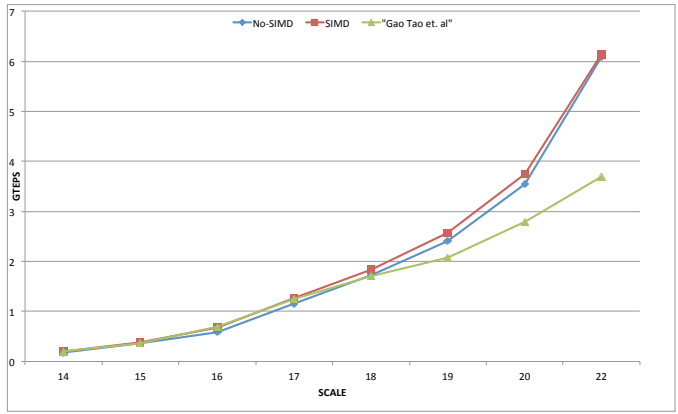

(a) $edgefactor = 16$

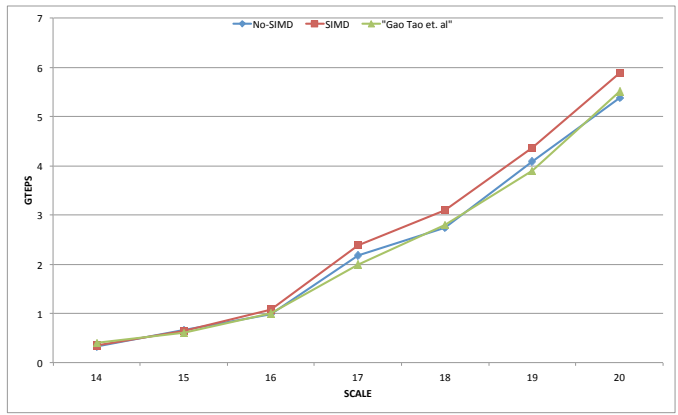

(b) $edgefactor = 32$

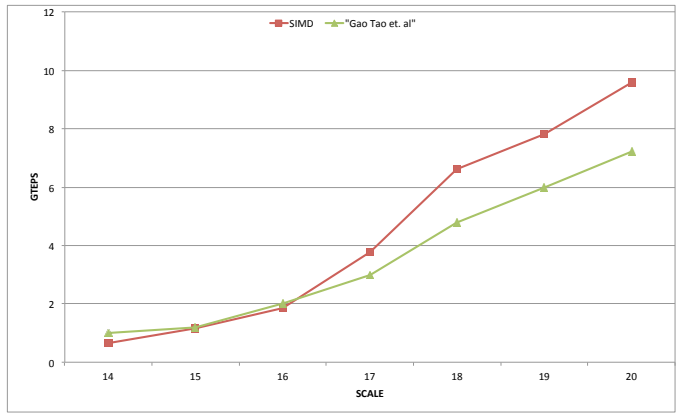

(c) $edgefactor = 64$

**Figure 3: Hybrid BFS performance results for the *SIMD-range* version and the state-of-the-art results presented by Gao *et al.* [5] for different graph sizes, using $SCALE$ from 14 up to 20 and edgefactor of 16, 32 and 64, generated by *Graph 500*.**

(2) In the three plots, there is a "hump" around $SCALE$ 17 and 18 for the graphs in Figure 3 a, b and c respectively. In particular the same "hump" can be observed in the implementation of the state-of-the-art Gao *et al.* [5] in Figure 3. This hump might be caused as an artifact of the Graph 500 during the graph generation, its analysis will part of the future work.

*SIMD versus the State-of-the-Art.* Figure 3 shows the results of the hybrid BFS algorithm using the *SIMD-range bottom-up* version and the results from the state-of-the-art vectorised hybrid BFS algorithm Gao *et al.* [5] for different graph sizes varying the $SCALE$ from 14 up to 22 and for *edgefactor* 16, 32 and 64 (a, b and c in the figure). For the three cases in Figure 3 (a, b and c), the performance of the SIMD version of this work (red line) is higher than the state-of-the-art *SIMD* hybrid BFS algorithm (green line). The state-of-the-art BFS algorithm is published by Gao *et al.* [5] and the results shown in Figure 3 (a, b and c) are the approximate values manually taken from the data presented in the paper. The approximate maximum speedup in each of the figures (a, b, c) is around 950 MTEPS (33%), 500 MTEPS (6.90%) and 1.50 GTEPS (22%).

Note that the source code of the Gao *et al.* implementation is not available, and thus it is not possible to make a direct analysis and comparison between our results and theirs.

The official *Graph 500* web-based ranking keeps an updated record of the performance of parallel computers. The performance is measured by running the BFS in different parallel architectures and according to the performance (number of TEPS) the computers are ranked. According to the most recently published list (June 2016) [1], an implementation of the hybrid BFS algorithm created by Golovina *et al.* [6] is executed on the same model of the Intel Xeon Phi used in this work (5110P), which is ranked in the 147th place in the list. This implementation is based on automatic vectorisation with two main optimisations: *loop unrolling* and *prefetching*, resulting in a performance of 1.80143 GTEPS for a graph size of $scale = 23$ and $edgefactor = 16$, using 60 cores. Despite the fact that in this work it is not possible to get the results for that graph size, due to the increment that the data structures utilised for vectorisation cause, our TEPS for a graph of $scale = 22$ and $edgefactor = 16$ is approximately 6 GTEPS, which is indicative of a better scaling than the results from Golovina *et al.* for the $scale = 23$. Though, the difference is that Golovina *et al.* used automatic vectorisation whereas in this work vectorisation is handled manually used by using intrinsic functions which implies the manual utilisation of the vector units.

## 9 CONCLUSIONS AND FUTURE WORK

The contributions of this work are, first, the development of an independent vectorisation of the hybrid BFS algorithm, which consists of both the SIMD *top-down* and the SIMD *bottom-up* that outperforms not only the state-of-the-art BFS algorithm developed by Gao *et al.* [1] but also overpasses Golovina's [6] BFS algorithm development listed in 147th position in the Graph 500 list. Second, a systematic analysis of the vectorised version of the bottom-up BFS algorithm performance based on hardware performance counters using PAPI library is presented, despite the fact that Gao *et al.* also implemented the vectorisation of the *bottom-up* algorithm, the performance they achieved was not fully understood because they left the impact of the vectorisation of the *bottom-up* BFS algorithm unclear. As a future work, the performance analysis of the vectorised version of the *bottom-up* BFS algorithm conducted in Section 7.2, using the PAPI library, raises some points to be investigated. For example, according to [9] the threshold for the CPI (cycles per instruction) in the Xeon Phi architecture should be investigated if it is higher than 4.0, which is in our results. Further analysis could help to understand more about the Xeon Phi architecture and the use of its features, including the vector unit and cache memory, which are crucial for graph traversals.

## ACKNOWLEDGMENT

This research was conducted with support from the UK Engineering and Physical Sciences Research Council (EPSRC) AnyScale Apps EP/L000725/1 and PAMELA EP/K008730/1. M. Paredes is funded by a National Council for Science and Technology of Mexico PhD Scholarship. Mikel Luján is supported by a Royal Society University Research Fellowship.

## REFERENCES

[1] 2016. The Graph 500 List. http://www.graph500.org. (2016).
[2] Scott Beamer, Krste Asanović, and David Patterson. 2012. Direction-optimizing Breadth-first Search. In *Proceedings of the International Conference on High Performance Computing, Networking, Storage and Analysis (SC '12)*. IEEE Computer Society Press, Los Alamitos, CA, USA, Article 12, 10 pages. http://dl.acm.org/citation.cfm?id=2388996.2389013
[3] Deepayan Chakrabarti, Yiping Zhan, and Christos Faloutsos. 2004. R-MAT: A Recursive Model for Graph Mining.. In *SDM*, Vol. 4. SIAM, 442–446.
[4] Barbara Chapman, Gabriele Jost, and Ruud van der Pas. 2007. *Using OpenMP: Portable Shared Memory Parallel Programming (Scientific and Engineering Computation)*. The MIT Press.
[5] Tao Gao, Yutong Lu, Baida Zhang, and Guang Suo. 2014. Using the Intel Many Integrated Core to accelerate graph traversal. *IJHPCA* 28, 3 (2014), 255–266. DOI:https://doi.org/10.1177/1094342014524240
[6] Elizaveta Andreevna Golovina, Aleksandr Sergeevich Semenov, and Aleksandr Sergeevich Frolov. 2014. Performance Evaluation of Breadth-First Search on Intel Xeon Phi. *Vychislitel'nye Metody i Programmirovanie* 15, 1 (2014), 49–58.
[7] John L. Hennessy and David A. Patterson. 2011. *Computer architecture: a quantitative approach*. Elsevier.
[8] Intel. 2012. *Optimization and Performance Tuning for Intel Xeon Phi Coprocessors*.
[9] Intel. 2012. *Optimization and Performance Tuning for Intel Xeon Phi Coprocessors, Part 2: Understanding and Using Hardware Events*.
[10] Intel Corporation. 2012. *A Guide to Vectorization with Intel® C++ Compilers*.
[11] Charles E Leiserson and Tao B Schardl. 2010. A work-efficient parallel breadth-first search algorithm (or how to cope with the nondeterminism of reducers). In *Proceedings of the twenty-second annual ACM symposium on Parallelism in algorithms and architectures*. ACM, 303–314.
[12] Jure Leskovec, Deepayan Chakrabarti, Jon Kleinberg, Christos Faloutsos, and Zoubin Ghahramani. 2010. Kronecker graphs: An approach to modeling networks. *Journal of Machine Learning Research* 11, Feb (2010), 985–1042.
[13] Richard C. Murphy, Kyle B. Wheeler, Brian W. Barrett, and James A. Ang. 2010. Introducing the graph 500. *Cray Users Group (CUG)* (2010).
[14] ICL University of Tennessee. 2016. *PAPI-C Programmer's Reference*.
[15] Mireya Paredes, Graham Riley, and Mikel Luján. 2016. Breadth First Search Vectorization on the Intel Xeon Phi. In *Proceedings of the ACM International Conference on Computing Frontiers 2016*.
[16] Rezaur Rahman. 2013. *Intel Xeon Phi Coprocessor Architecture and Tools: The Guide for Application Developers* (1st ed.). Apress, Berkely, CA, USA.
[17] Erik Saule and Ümit V Çatalyürek. 2012. An early evaluation of the scalability of graph algorithms on the Intel MIC architecture. In *Parallel and Distributed Processing Symposium Workshops & PhD Forum (IPDPSW), 2012 IEEE 26th International*. IEEE, 1629–1639.
[18] Milan Stanic, Oscar Palomar, Ivan Ratkovic, Milovan Duric, Osman Unsal, Adrian Cristal, and Mateo Valero. 2014. Evaluation of vectorization potential of graph500 on Intel's Xeon Phi. In *High Performance Computing & Simulation (HPCS), 2014 International Conference on*. IEEE, 47–54.
[19] Gao Tao, Lu Yutong, and Suo Guang. 2013. Using MIC to accelerate a typical data-intensive application: the breadth-first search. In *Parallel and Distributed Processing Symposium Workshops & PhD Forum (IPDPSW), 2013 IEEE 27th International*. IEEE, 1117–1125.
[20] Chenxu Wang, Yutong Lu, Baida Zhang, Tao Gao, and Peng Cheng. 2015. An optimized BFS algorithm: A path to load balancing in MIC. In *Computer and Communications (ICCC), 2015 IEEE International Conference on*. IEEE, 199–206.
[21] Jeremiah Willcock. 2012. Graph 500 Benchmark. (nov 2012). Presentation at Super Computing 2012.

---

[1] Gao *et al.* [5] results are not listed in the Graph 500 list.